\documentclass[twocolumn,showpacs,amsmath,amssymb]{revtex4}

\usepackage{graphicx} 
\usepackage{dcolumn}
\usepackage{bm}
\usepackage{ifpdf}

\begin{document}

\title{Generalized sine-Gordon solitons}

\author{C. dos Santos$^1$, D. Rubiera-Garcia$^2$ }

\affiliation{$^1$Centro de F\'{\i}sica e Departamento de F\'{\i}sica e Astronomia, Faculdade de Ci\^{e}ncias da Universidade do Porto, 4169-007 Porto, Portugal.\\
$^2$Departamento de F\'{\i}sica, Universidad de Oviedo, Avenida Calvo Sotelo 18, 33007 Oviedo, Asturias, Spain.}

\date{\today}

\begin{abstract}

In this paper we construct analytical self-dual soliton solutions in (1+1) dimensions for two families of models which can be seen as generalizations of the sine-Gordon system but where the kinetic term is non-canonical. For that purpose we use a projection method applied to the Sine-Gordon soliton. We focus our attention on the wall and lump-like soliton solutions of these k-field models. These solutions and their potentials reduce to those of the Klein-Gordon kink and the standard lump for the case of canonical kinetic term. As we increase the non-linearity on the kinetic term the corresponding potentials get modified and the nature of the soliton may change, in particular, undergoing a topology modification. The procedure constructed here is shown to be a sort of generalization of the deformation method for a specific class of k-field models.

\end{abstract}

\maketitle
\section{Introduction}

K-fields are field theories were the kinetic term is non-canonical and whose soliton solutions have been intensively studied due to their applications in strong interaction physics \cite{strong}, topological defects \cite{defects,defect2} and cosmology \cite{cosmology} with the result that their properties can be quite different from the standard canonical ones. For example these solitons can have a compact support \cite{Adametal}.

The aim of this paper is to present a method to construct \emph{analytical} soliton solutions in $(1+1)$ dimensions in k-field theories and relate it with the deformation method developed by Bazeia et al. and originally introduced in Ref.\cite{Bazeia2}. Such soliton solutions may be of wall or lump-like type. We take the context of the sine-Gordon model \cite{Rubinstein}, as one of the most interesting solitons in  integrable $(1+1)$ dimensional systems.

The deformation method, used in a wide range of contexts, has also been employed in the sine-Gordon one and its modifications where the scalar potential $V(\phi)$ is given by polynomial interactions obtained from truncations of the sine-Gordon one \cite{lohe79} allowing then to get new families of sine-Gordon models \cite{Bazeia1}.

Sine-Gordon solitons have physically attractive properties, as they can be used to describe many physical phenomena, such as the Josephson junctions \cite{Scott} and systems with one-dimensional dislocations \cite{Lamb}, but also mathematical ones such as the description of spaces with constant negative curvature \cite{Eisenhart} or simply being objects of interest by their own on integrable and conformal field theories \cite{Babelon}.

Moreover solitons are of current interest due to its applications to supergravity \cite{Cvetic}, brane cosmology \cite{Davis}, but also to other areas of nonlinear science \cite{Collins}, in particular in the study of macromolecules such as the DNA \cite{Bryant}.

In several of the above areas the deformation method has been shown to be a useful tool for the analysis of diverse problems. The method developed in this paper extends that one to a class of k-fields defined as powers of the canonical kinetic term, using as the starting point the sine-Gordon model. This kind of choice in the kinetic term for electromagnetic fields has been considered in Ref.\cite{Hassaine}. By doing this we shall obtain several families of k-field models with kink and lump-like solutions, which are of current interest to high-energy physics; for example, in the context of their embedding in higher dimensions as they may play a role in the braneworld context \cite{Gomes08}, and as cosmological domain walls \cite{Avelino}.

This paper is organized as follows: in Sec II we briefly review the sine-Gordon model and its cosine-Gordon counterpart. In Sec. III we present the generic formalism for k-field theories and discuss the necessary conditions to have self-dual solutions. In Sec. IV we present several examples of k-field models, obtain analytical soliton solutions and study their physical properties. In Sec. V we compare our results with those obtained by using the deformation method and we conclude in section VI by drawing some conclusions and future perspectives.

\section{Sine-Gordon system}

This section is divided into two parts: in the first one we briefly review static soliton solutions in (1+1) dimensions for the sine-Gordon model \cite{Rubinstein}. In the second part we employ the same procedure but now for analysis of the cosine-Gordon model.

\subsection{Sine-Gordon model}

The ($1+1$)-dimensional lagrangian density for the sine-Gordon model \cite{Rubinstein} is given by
\begin{equation}
L=\frac{1}{2}\partial_{\mu}\phi\partial^{\mu}\phi-\frac{\alpha}{\beta^2}(1-\cos[\beta\phi]) \label{SG},
\end{equation}
where $\phi$ is a single real scalar field and $\alpha$ and $\beta$ are two positive parameters. By convenience we take $\alpha=1$ and $\beta=2$ which brings that the scalar potential writes
\begin{equation}
V(\phi)=\frac{1}{2}\,\sin^2[\phi]\label{potential1}.
\end{equation}
For further reference we also write
\begin{equation}
X=\frac{1}{2}\partial_{\mu}\phi\partial^{\mu}\phi, \label{dalembert}
\end{equation}
for the standard canonical kinetic term.

For static fields the Euler Lagrange equations of motion are
\begin{equation}
\phi^{\prime \prime} = \frac{1}{2}\sin[2\phi],
\end{equation}
which for the Bogomon'nyi-Prasad-Sommerfeld (BPS) states \cite{Prasad} becomes of first order and is given by
the relation
\begin{equation}
\phi'=\sin[\phi].
\end{equation}
The kink solution is well known \cite{Ablowitz} and is given by
\begin{equation}
\phi_{0}(x)=2 \arctan[e^{(x-x_0)}] \label{SGsol},
\end{equation}
which is located at $x=x_0$ and satisfies the boundary conditions
$\phi(x=-\infty)=0$ and $\phi(x=+\infty)=\pi$.

Let us note that the standard Klein-Gordon kink solution, i.e., the soliton solution for the standard Higgs potential
$\frac{1}{2}(1-\phi^2)^2$, which is given by $\tanh[x-x_0]$ may be written in terms of the sine-Gordon solution, $\phi_0$, as
\begin{equation}
\phi_1 = -\cos[\phi_0].
\end{equation}
This means that the standard kink soliton solution may be seen as a projection of the field $\phi_0$ (which thus can be seen as an $``$angle$"$) over the real axis.

In the same way the standard lump, soliton solution of the potential $\frac{1}{2}(1-\phi^2)\phi^2$, which is given by
$\cosh[x-x_0]^{-1}$, can be seen as a projection of the sine-Gordon field $\phi_0$ but now over the imaginary axis
\begin{equation}
\phi_2 = \sin[\phi_0].
\end{equation}

We now note that the sine-Gordon potential, i.e.,
\begin{equation}
V(\phi)=\frac{\alpha}{\beta^2}(1-\cos[\beta\phi]) \label{potential},
\end{equation}
has a infinite number of vacua given by the solutions $\phi_0= 2\pi n / \beta$. Consequently the lagrangian (\ref{SG}) with this potential is invariant under shifts of the field of the form $\phi_0 \rightarrow \phi_0+ 2\pi n / \beta$ leading to different topological sectors labeled by the charge $Q=\frac{\beta}{2\pi}(\phi_0(+\infty)-\phi_0(-\infty))$.

Let us stress that besides the static kink soliton solution, the sine-Gordon model also admits anti-kink, breather, wobbles and kink-breather solutions \cite{Ferreira}.

\subsection{Cosine-Gordon model}

Let us now perform a field shift in the potential given in (\ref{potential1}) to take
\begin{equation}
\overline{\phi} = {\phi}-\frac{\pi}{2}, \label{shiftpiover2}
\end{equation}
bringing that the potential now becomes
\begin{equation}
 \overline{V}( \overline{\phi})=\frac{1}{2} \cos^2[ \overline{\phi}],
\end{equation}
which shall be called, by convenience, the ``cosine-Gordon" model. Its static BPS kink is given by
\begin{equation}
\overline{\phi}_0(x)=2\arctan[\tanh[\frac{x-\overline{x}_0}{2}]], \label{CGsol}
\end{equation}
which is the solution of the first order equation of motion
\begin{equation}
\overline{\phi}'=\cos[\overline{\phi}],
\end{equation}
with the boundary conditions $\overline{\phi}(x=-\infty)=-\frac{\pi}{2}$ and
$\overline{\phi}(x=+\infty)=\frac{\pi}{2}$.

One gets that now, when applying the $``$projection$"$ method underlined above, the standard Klein-Gordon kink and the lump soliton solution considered above, i.e., $\phi_W$ and $\phi_L$ are, respectively, given by
\begin{equation}
\phi_W=\sin(\overline{\phi}); \phi_L=\cos(\overline{\phi}).
\end{equation}

The reason to consider this ``cosine-Gordon" model, as shall become clear later, is the fact that the projected models constructed on Sec. IV, i.e., models of family A and B, \emph{do not retain the original symmetry of the sine-Gordon model}. Therefore, the projections applied to the sine-Gordon or to the cosine-Gordon model will be, in general, different, as shall be seen at once.

\section{Generic formalism for k-field models}

Let us now introduce the generic formalism for k-field models with non-canonical kinetic term $F(X)$ and potential
$V(\phi)$ in $(1+1)$ dimensions. Its action is given by

\begin{equation}
S=\int d^2 x L= \int d^2 x \left(F(X)-V(\phi)\right), \label{action}
\end{equation}
where $F(X)$ is an arbitrary function of the D'Alambert lagrangian defined in Eq.(\ref{dalembert}). The associated energy-momentum tensor is obtained as

\begin{equation}
T^{\mu\nu}=F_X \partial^{\mu}\phi \partial^{\nu} \phi -\eta^{\mu\nu}L.
\end{equation}
Frow now one we shall only consider static configurations $\phi=\phi(x)$, for which the non-vanishing components of the energy-momentum tensor are given by (we are defining $F_X\equiv \frac{dF}{dX}$)

\begin{eqnarray}
T^{00}&=&\varepsilon=-L=-F(X)+V(\phi) \nonumber \\
T^{11}&=&-p=F(X)-2XF_X, \label{em}
\end{eqnarray}
where $\varepsilon$ and $p$ are the energy density and the pressure, respectively. Note that the positive definiteness of the energy density implies $F_X>0$.

The equations of motion for the action (\ref{action}) are obtained as
\begin{equation}
\partial_{\mu} (F_X \partial^{\mu}\phi)+\frac{\partial  V}{\partial \phi}=0. \label{eom}
\end{equation}
These equations, for static solutions, have one first integral given by $F(X)-2XF_X-V=D$, where the integration constant $D$ turns out to be the pressure $p=D$. We shall only deal with BPS states \cite{Prasad}, which implies the condition $p=0$. Moreover, this condition is also necessary for stability under Derrick's scaling, as shown in Ref.\cite{defect2}. As the pressureless of the source makes the constant $D$ to vanish the first integral for our problem is given by
\begin{equation}
F(X)-2XF_X=V \label{FI}.
\end{equation}

In the next section we shall use the $``$projection$"$ method underlined in section II to construct two families of k-field models, $A$ and $B$. The family $A$ has two models, \emph{$A1$} and \emph{$A2$}, whose soliton solutions are analytically obtained as projections of the $``$angle$"$ $\overline{\phi}_0 /a$ along the imaginary/real axis, respectively, with $a$ an integer positive number which is precisely the power of the non-canonical kinetic term, i.e. $X^a$. Family $B$ has also two models, \emph{$B1$} and \emph{$B2$} whose solutions are also obtained as projections but now of the $``$angle$"$ ${\phi}_0 /a$ along the imaginary/real axis.

\section{Family $A$ of k-field models}

This section is divided into two parts depending on the value of $a$ which is taken. In the first one we take $a=2$ to illustrate the general procedure while in the second one we extend the former results and take $a>2$.

\subsection{Family $A$ of k-field models with $X^2$}

Let us consider two models: the model $L$ and the model $W$ which support lump-like and wall soliton solutions respectively, as shown below,
and whose lagrangians are given by

\begin{equation}
L_{L}=X |X|-\frac{3}{4}\sin^{4}(\overline{\phi}_{0}/2)\cos^{4}(\overline{\phi}_{0}),
 \label{modelL2}
\end{equation}
and
\begin{equation}
L_{W}=X |X|-\frac{3}{4}\cos^{4}(\overline{\phi}_{0}/2)\cos^{4}(\overline{\phi}_{0}),
 \label{modelW2}
\end{equation}

The first model, (\ref{modelL2}), admits the analytical solution
\begin{equation}
\phi_L(x)=2 \cos\left[\frac{\overline{\phi}_{0}(x)}{2}\right], \label{lump}
\end{equation}
\begin{figure}[h]
\begin{center}
\includegraphics[width=8cm,height=4.5cm]{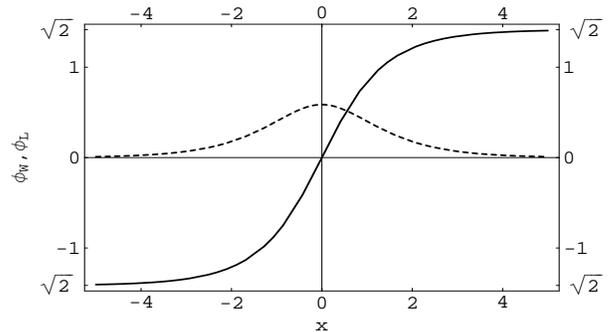}
\caption{\label{fig:1} Profile of the lump-like solution (dashed line) of model (\ref{modelL2}) and the wall (solid) of model
(\ref{modelW2}) versus the distance $x$. The soliton is located at $x=0$.}
\end{center}
\end{figure}
where $\overline{\phi}_{0}$ is given in Eq.(\ref{CGsol}). This is a lump-like solution, whose profile is shown in figure 1 (dashed line). The energy density associated to this field is given by $\varepsilon=4X^2=\frac{4}{3}V$ (see figure 2) where the last equality is a simple consequence of Eq.(\ref{FI}). Note that the lump-like solution (\ref{lump}) can be seen as a $``$projection$"$ over the real axis of the $``$angle$"$ $\overline{\phi}_{0}/2$.

The second model, (\ref{modelW2}), admits the analytical solution
\begin{equation}
\phi_W(x)= 2 \sin\left[\frac{\overline{\phi}_{0}(x)}{2}\right]. \label{wall}
\end{equation}
Note that the field that we have constructed in (\ref{wall}) is a wall type one (see figure 1) with a topological charge $Q=\frac{1}{2\sqrt{2}}\int_{-\infty}^{+\infty} \phi^{'}_W(x)dx$, and whose energy density is displayed in figure 2 (solid line).
\begin{figure}[h]
\begin{center}
\includegraphics[width=8cm,height=4.5cm]{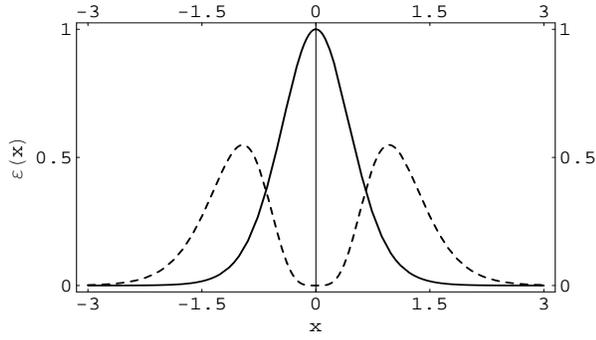}
\caption{\label{fig:2} Energy density versus the distance $x$ for the wall solution (\ref{wall}) (solid line) and the
lump-like solution (\ref{lump}) (dashed). The lump density has been increased by a factor of $100$
in order to be plotted in the same figure as the wall one.}
\end{center}
\end{figure}

We now note that the two models, (\ref{modelL2}) and (\ref{modelW2}), can be merged into a single one whose lagrangian density is simply given by

\begin{equation}
L=X |X|-\frac{3}{4}\left(1-\frac{\phi^2}{4}\right)^2\left(-1+\frac{\phi^2}{2}\right)^4. \label{x2-model}
\end{equation}

Its equations of motion for BPS states, given by (\ref{eom}), are obtained as

\begin{equation}
-3X |X|=V \rightarrow \phi^{'4}=\left(1-\frac{\phi^2}{4}\right)^2\left(-1+\frac{\phi^2}{2}\right)^4.
\end{equation}

To see that the potential in Eq.(\ref{x2-model}) effectively gives the solution $\phi_W$, starting from ($\ref{wall}$) one can write $X=-\frac{1}{2}\phi_W^{'2}(x)=-\frac{1}{2} \cos^2\left[\overline{\phi}_0/2\right] \cos^2[\overline{\phi}_{0}]$ and by transforming the $\cos$ terms into $\sin$ ones through trigonometric relations, and rewriting the final expression in terms of the field (\ref{wall}) one arrives to $X=-\frac{1}{2} \left[\left(1-\frac{\phi^2}{2}\right)^2\left(1-\frac{\phi^2}{4}\right)\right]$, and to the associated potential $V(\phi)=-3X |X|=\frac{3}{4}\left[\left(1-\frac{\phi^2}{2}\right)^4\left(1-\frac{\phi^2}{4}\right)^2\right]$, which is precisely the one of (\ref{x2-model}). A similar reasoning can be applied to the field $\phi_L$ of Eq.(\ref{lump}), ending up in the lagrangian density (\ref{x2-model}) as well.

\subsection{Family $A$ of k-field models with $X^a$ where $a>2$}

The previous model is a particular case of a more general family of models defined by lagrangian densities
\begin{equation}
L_{L}=X |X|^{a-1}-\frac{(2a-1)}{2^a}\sin^{2a}(\overline{\phi}_{0}/a)\cos^{2a}(\overline{\phi}_{0}),
 \label{modelLa}
\end{equation}
and
\begin{equation}
L_{W}=X |X|^{a-1}-\frac{(2a-1)}{2^a}\cos^{2a}(\overline{\phi}_{0}/a)\cos^{2a}(\overline{\phi}_{0}),
 \label{modelWa}
\end{equation}
which admit respectively the solutions
\begin{equation}
\phi_L(x)=a \cos\left[\frac{\overline{\phi}_{0}(x)}{a}\right] ; \phi_W(x)=a \sin\left[\frac{\overline{\phi}_{0}(x)}{a}\right], \label{generalized-lw}
\end{equation}
where we keep the notation $L$ and $W$ of the previous section. Note that the positivity of the energy (\ref{em}) imposes the condition $a>1/2$ in the models above. Moreover the hyperbolicity condition for perturbations \cite{hiperbolicity}
\begin{equation}
\frac{2XF_{XX}+F_X}{F_X}>0,
\end{equation}
is satisfied for $a>1/2$ which means that for the case considered here small perturbations over the background do not grow exponentially in time, ensuring the stability of the solutions.

\subsubsection{Properties of the potentials $V_L$ and $V_W$}

We now proceed further by studying the properties of the potentials $V_L$ and $V_W$ in order to argue the existence of soliton solutions.

First we note that when $a=1$ the solutions in (\ref{generalized-lw}) and the potentials in (\ref{modelLa}),(\ref{modelWa}) reduce to those of the standard lump and Klein-Gordon kink, respectively.

On the other hand for $a=2$ these potentials turn out to be the same and moreover equal to the one in (\ref{x2-model}) and thus we recover the above solutions (\ref{lump}) and (\ref{wall}), as expected. For greater values of $a$, one gets that when $a$ is odd the potentials $V_L$ and $V_W$ are different (as we already saw for the particular case $a=1$), while for $a$ even they coincide when written in terms of the fields $\phi_L$, $\phi_W$. For example, for $a=4$ we get

\begin{equation}
V_L=V_W=\frac{7}{16}\Big[1-\Big(\frac{\phi_{L,W}}{4}\Big)^2\Big]^4
\Big [-1+2\Big(1-2\Big(\frac{\phi_{L,W}}{4}\Big)^2\Big)^2\Big]^8.
\end{equation}

Also, for both potentials, as $a$ increases the number of vacua also increases, adding more topological sectors. In order to see the evolution on the number of these sectors with $a$, let us carefully analyze the vacua structure of these potentials.

$V_W$ is always positive and its zeroes are given by

\begin{equation}
\overline{Z}_n=a \sin\left[\left(n-\frac{1}{2}\right)\frac{\pi}{a}\right],
\end{equation}
where $n\epsilon \mathbb{N}$. The count of different zeros runs from $n=1$ to $n=a/2+1$ ($a$ even) or to $n=(a+1)/2$ ($a$ odd). Explicitly the results are

\begin{equation}
\{\overline{Z}_1=a,\overline{Z}_2,\cdot \cdot \cdot,\overline{Z}_{\frac{a}{2}+1},-\overline{Z}_{\frac{a}{2}+1},\cdot \cdot \cdot,-\overline{Z}_2,-\overline{Z}_1=-a \},
\end{equation}
for $a$ even and

\begin{equation}
\{\overline{Z}_1=a,\overline{Z}_2,\cdot \cdot \cdot,\overline{Z}_{\frac{a+1}{2}},-\overline{Z}_{\frac{a+1}{2}},\cdot \cdot \cdot,-\overline{Z}_2,-\overline{Z}_1=-a \},
\end{equation}
for $a$ odd. The structure of zeroes is therefore symmetric in both cases.

Concerning the potential $V_L$, its zeroes are given by

\begin{equation}
\overline{Z}^{'}_n=a \cos\left[\left(n-\frac{1}{2}\right)\frac{\pi}{a}\right],
\end{equation}
and, explicitly, for $a$ even the zeroes are obtained as

\begin{equation}
\{\overline{Z}^{'}_1=a,\overline{Z}^{'}_2,\cdot \cdot \cdot,\overline{Z}^{'}_{\frac{a}{2}+1},-\overline{Z}^{'}_{\frac{a}{2}+1},\cdot \cdot \cdot,-\overline{Z}^{'}_2,-\overline{Z}^{'}_1=-a \},
\end{equation}
while for $a$ odd we have

\begin{equation}
\{a,\overline{Z}^{'}_1,\overline{Z}^{'}_2,\cdot \cdot \cdot,\overline{Z}^{'}_{\frac{a-1}{2}},\overline{Z}^{'}_{\frac{a+1}{2}}=0,-\overline{Z}^{'}_{\frac{a-1}{2}},\cdot \cdot \cdot,-\overline{Z}^{'}_2,-\overline{Z}^{'}_1,-a \},
\end{equation}
The potential $V_L$ is always positive when $a$ is even while for $a$ odd it becomes negative beyond the farther zero ($\pm a$) of the potential and moreover there is an additional central vacuum as compared with the other potentials. The height of the local maximums in between each pair of vacua decreases with $a$ for both $V_L$ and $V_W$. In figure 3 we have plotted this kind of behaviour for $a$ odd ($a=3$) for both $V_L$ and $V_W$, as an illustrative example of these families of potentials. For $a$ even these potentials are equal, as already stated, and with a similar behaviour as the one for walls of $a$ odd (solid line in Fig.3).

\begin{figure}[h]
\begin{center}
\includegraphics[width=8cm,height=4.5cm]{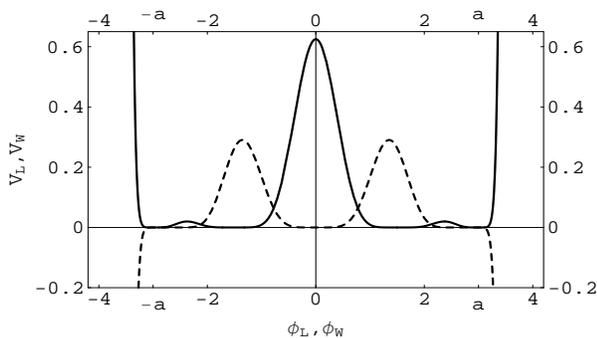}
\caption{\label{fig:3} The potentials $V_W$ versus $\phi_W$ (solid line) and
$V_L$ versus $\phi_L$ (dashed) given in  (\ref{modelLa}) and (\ref{modelWa}) for the case $a=3$.}
\end{center}
\end{figure}

\subsubsection{Evolution of $\phi_W$ and $\phi_L$ with $a$}

Let us now analyze the evolution of the soliton solutions $\phi_W$ and $\phi_L$ with the parameter $a$. We will illustrate it by making use of some plots combined with the analysis of the topological charge.

We begin with $\phi_W$. In figure 4 we plot its evolution when $a$ increases. As $a=1$ it gives the Klein-Gordon kink ($a=1$), with boundary conditions $\phi(\pm \infty)=\pm 1$. As $a$ increases the height of the wall increases and in the limit of large $a$ this family $A$ of solutions approaches the cosine-Gordon soliton, $\overline{\phi}_{0}$, given in Eq.(\ref{CGsol}), with boundary values $\overline{\phi}_0(\pm \infty)=\pm \pi/2$. This can be easily seen by looking at the evolution of the topological charge: by introducing a topological current as

\begin{equation}
J^{\mu}_W=\frac{1}{2}\varepsilon^{\mu\nu}\partial_{\nu}\phi_W,
\end{equation}
we get for this case the (non-normalized) topological charge
\begin{eqnarray}
Q_W&=&\int_{-\infty}^{+\infty} dx J^0_W=\frac{1}{2}[\phi_W(+\infty)-\phi_W(-\infty)]\nonumber\\
&=&\frac{a}{2}(\sin[\overline{\phi}_0(+\infty)/a]-\sin[\overline{\phi}_0(-\infty)/a]) = \\ &=&a\sin\left[\frac{\pi}{2a}\right], \nonumber
\end{eqnarray}
\begin{figure}[h]
\begin{center}
\includegraphics[width=8cm,height=4.5cm]{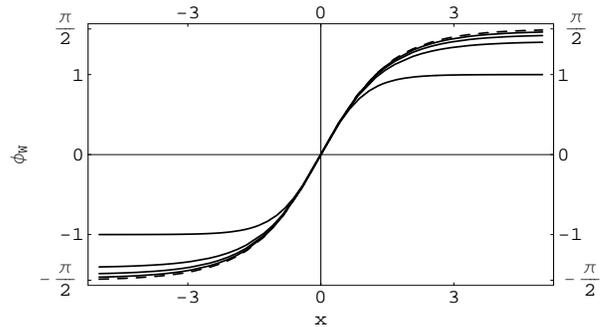}
\caption{\label{fig:3} Evolution with $a$ ($=1,2,3,4,5$) of the wall profile, $\phi_W$ versus the distance $x$ for the family of models (\ref{modelWa}).
The dashed line corresponds to the cosine-Gordon field profile, $\overline{\phi}_{0}(x)$, to which $\phi_W$ converges in the large $a$ limit.
All the solitons are located at $x=0$.}
\end{center}
\end{figure}
which in particular gives $Q_W(a=1)=1$ and $Q_W(a\rightarrow \infty) \rightarrow \pi/2$, thus connecting through a monotonic function of $a$
the Klein-Gordon kink and the cosine-Gordon field. The non-vanishing of this topological charge indicates the presence of field configurations
with a non-trivial topology, and whose presence guarantees the existence and stability of the wall.

\begin{figure}[h]
\begin{center}
\includegraphics[width=8cm,height=4.5cm]{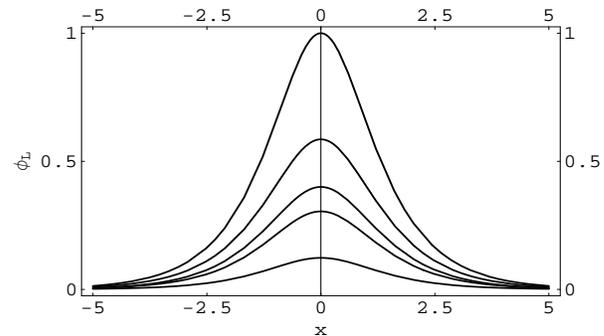}
\caption{\label{fig:3} From top to bottom, evolution with $a$ ($=1,2,3,4,10$) of the lump profile, $\phi_L$, versus the distance $x$
for the family of models (\ref{modelLa}). All the solutions are located at $x=0$ by applying appropriate shifts.}
\end{center}
\end{figure}
We now analyze the evolution of $\phi_L$. From Fig.5 we see that the height of the lump decreases as $a$ increases as compared with the case $a=1$ (for which $\phi_L(x=0) = 1$), becoming more spread in space and disappearing in the limit of $a \rightarrow \infty$. Calculating the topological charge associated to any of these solutions as in the previous case one gets
\begin{eqnarray}
Q_L&=&\frac{1}{2}[\phi_L(+\infty)-\phi_L(-\infty)]\nonumber\\
&=&\frac{a}{2}(\cos[\overline{\phi}_0(+\infty)/a]-\cos[\overline{\phi}_0(-\infty)/a])
\end{eqnarray}
which is identically zero for any value of $a$, as expected.

\subsubsection{The energy density}

The energy density associated to models (\ref{modelLa})-(\ref{modelWa}) is obtained as

\begin{eqnarray}
\varepsilon_W(a)&=&-F(X)+V_W(\phi)=-2XF_X=-2aX |X|^{a-1}= \nonumber \\ &=&\frac{2a}{2a-1}V_W=\frac{2a}{2^a}\cos^{2a}[\overline{\phi}_{0}/a]\cos^{2a}[\overline{\phi}_{0}] \label{energy2}\\
\varepsilon_L(a)&=&-F(X)+V_L(\phi)=-2XF_X=-2aX |X|^{a-1}= \nonumber \\ &=&\frac{2a}{2a-1}V_L=\frac{2a}{2^a}\sin^{2a}[\overline{\phi}_{0}/a]\cos^{2a}[\overline{\phi}_{0}] \label{energy}
\end{eqnarray}
for walls and lumps, respectively. Its evolution with $a$ for the case of walls is shown in figure 6, where we see that the energy density becomes more localized as we increase $a$.

\begin{figure}[h]
\begin{center}
\includegraphics[width=8cm,height=4.5cm]{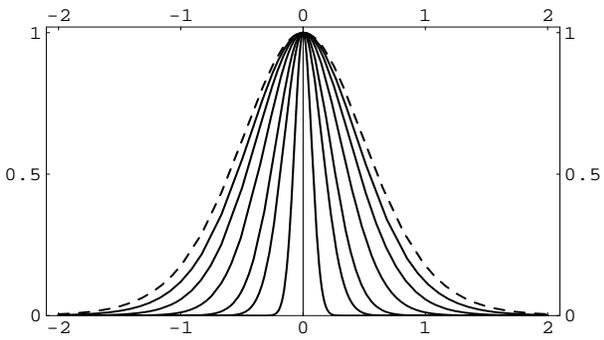}, \caption{\label{fig:3} From the exterior (dashed curve) to the interior, evolution of the factor $\cos^{2a}[\overline{\phi}_{0}/a]\cos^{2a}[\overline{\phi}_{0}]$ in Eq.(\ref{energy2}) with $a$ ($=2,3,5,10,20,100$) for the wall solution and versus the
distance $x$. All the solitons are located at $x=0$. The dashed line corresponds to the
canonical kinetic term case ($a=1$).}
\end{center}
\end{figure}

\subsection{Family $B$ of k-field models with $X^a$}

Let us consider now another family of models with lagrangian densities

\begin{equation}
L_{1}=X |X|^{a-1}-\frac{(2a-1)}{2^a}\sin^{2a}[\phi_0/a]\sin^{2a}[\phi_0],
 \label{model1}
\end{equation}
and
\begin{equation}
L_{2}=X |X|^{a-1}-\frac{(2a-1)}{2^a}\cos^{2a}[\phi_0/a]\sin^{2a}[\phi_0],
 \label{model2}
\end{equation}
which admit analytical solutions constructed as projections over the real and imaginary axes of the solutions of the sine-Gordon system given in Eq.(\ref{SGsol}), i.e.

\begin{equation}
\phi_1(x)=-a \cos\left[\frac{\phi_0(x)}{a}\right] ; \phi_2(x)=a \sin\left[\frac{\phi_0(x)}{a}\right]. \label{solutions2}
\end{equation}

\subsubsection{Properties of the potentials $V_1$ and $V_2$}

From the expressions above it follows that both potentials coincide when $a$ is even, and differ when $a$ is odd, as for the family $A$. For example, for the case $a=2$, expanding the above expressions we get

\begin{equation}
V_1(a=2)=V_2(a=2)=\frac{3}{4}\left(1-\frac{\phi_{1,2}^2}{4}\right)^4\phi_{1,2}^4,
\end{equation}
which shows their equal analytical expression in terms of the fields $\phi_1$ and $\phi_2$. Note that in this potential the point $\phi_1=0$ is now a minimum, instead of a maximum, as opposed to the previous family of models. Again this is a general result for $a$ even: in this case there are $(\frac{a}{2})$ symmetric vacua and an additional central vacuum for both potentials, which are always positive. For $a$ odd there are $(\frac{a+1}{2})$ symmetric vacua for both $V_1$ and $V_2$. However, there is now an additional central vacuum for $V_2$, and this potential becomes negative beyond the farther vacuum, while $V_1$ is always positive and reaches a maximum at the center. In all cases the height of each successive maximum of the potential decreases as we move away from $\phi=0$. The zeroes are obtained, for $V_1$, as

\begin{equation}
Z_n=a \cos\left[n\frac{\pi}{a}\right],
\end{equation}
and are given, explicitly, by
\begin{equation}
\{Z_1=a,Z_2,\cdot \cdot \cdot,Z_{\frac{a}{2}},Z_{\frac{a}{2}+1}=0,-Z_{\frac{a}{2}},\cdot \cdot \cdot,-Z_2,-Z_1=-a \},
\end{equation}
for $a$ even and

\begin{equation}
\{Z_1=a,Z_2,\cdot \cdot \cdot,Z_{\frac{a+1}{2}},-Z_{\frac{a+1}{2}},\cdot \cdot \cdot,-Z_2,-Z_1=-a \},
\end{equation}
for $a$ odd. And for $V_2$ the zeroes

\begin{equation}
Z^{'}_n=a \sin\left[n\frac{\pi}{a}\right],
\end{equation}
are obtained as

\begin{equation}
\{Z^{'}_1=a,Z^{'}_2,\cdot \cdot \cdot,Z^{'}_{\frac{a}{2}},Z^{'}_{\frac{a}{2}+1}=0,-Z^{'}_{\frac{a}{2}},\cdot \cdot \cdot,-Z_2,-Z_1=-a \},
\end{equation}
for $a$ even and

\begin{equation}
\{a,Z^{'}_1,Z^{'}_2,\cdot \cdot \cdot,Z^{'}_{\frac{a-1}{2}},Z^{'}_{\frac{a+1}{2}}=0,-Z^{'}_{\frac{a-1}{2}},\cdot \cdot \cdot,-Z^{'}_2,-Z^{'}_1=-a \}.
\end{equation}
for $a$ odd. Note that in the case $a=1$ we recover, for $V_1$, the Mexican hat potential of the $\phi^4$-model while for $V_2$ we recover the potential of the standard lump, a similar situation as for the potentials $V_W$ and $V_L$ of the previous family $A$. This is indeed a general conclusion for $a$ odd: the potentials $V_1$ and $V_2$ coincide, as functions of their arguments, with $V_W$ and $V_L$, respectively. The typical behaviour of $V_1$ and $V_2$ in this case is similar as for the first family ($A$) and figure 3 above is again illustrative of this generic behaviour. This is not so when $a$ is even as the potential has now a minimum at $\phi=0$ and it is positive everywhere, thus resembling neither $V_W$ nor $V_L$.

\subsubsection{Evolution of $\phi_1$ and $\phi_2$ with $a$}

For $a=1$ $\phi_1$ describes the Klein-Gordon kink while $\phi_2$ corresponds to the standard lump potential, as expected from the analysis of the potentials in this case. This is in agreement with the fact that through a $\pi/2$ rotation between the imaginary and real axis one changes the boundary conditions of the soliton solutions in such as way that the topological wall solution may be changed into a non-topological lump-like one and viceversa. Now, as we increase $a$ we can follow the evolution of $\phi_1$,
which leads to wall-like solutions which become thicker with $a$ until they disappear in the limit $a \rightarrow \infty$. This can be seen through the evolution of the (non-normalized) topological charge, written as

\begin{eqnarray}
Q_1(a)&=& \int_{-\infty}^{+\infty} \phi_1^{'}(x)dx=\frac{a}{2}[\phi_1(+\infty)-\phi_1(-\infty)]= \nonumber \\ &=&  -\frac{a}{2}(\cos[\phi_0(+\infty)/a]-\cos[\phi_0(-\infty)/a]) \\ &=& \frac{a}{2}\left[1-\cos\left[\frac{\pi}{a}\right]\right] \nonumber
\end{eqnarray}
which gives $Q_1(a=1)=1=Q_{KG}$ and $Q_1(a \rightarrow \infty) \rightarrow 0$, as expected. Now if we look at the evolution of $\phi_2$ with $a$, it can shown that its non-normalized charge is given by

\begin{equation}
Q_2(a)=\frac{a}{2}(\sin[\phi(+\infty)/a]-\sin[\phi(-\infty)/a])=\frac{a}{2}\sin\left[\frac{\pi}{a}\right]
\end{equation}
which vanishes for $a=1$, as is expected for the lump, but for $a=2$ takes the value $Q_2=2$, undergoing a topology change from a lump-like solution to a wall one. This evolution can be followed through figure 7, where the soliton profile is plotted. There the solution begins being a lump-like, transforms into a wall and in the limit $a \rightarrow \infty$ the family of solutions approaches to the sine-Gordon kink since $Q(a \rightarrow \infty) \rightarrow \pi=Q_{SG}$ when both charges are non-normalized.
\begin{figure}[h]
\begin{center}
\includegraphics[width=8cm,height=4.5cm]{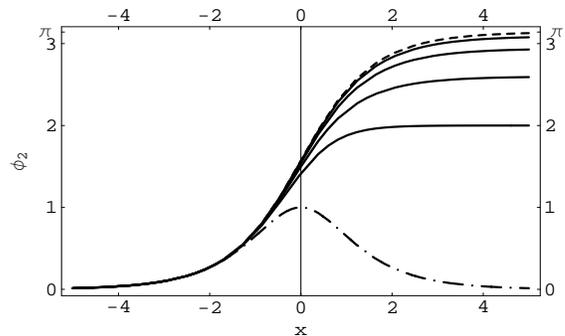}
\caption{\label{fig:7} Soliton profile of $\phi_2$ versus the distance $x$ for different values of (from bottom to top) $a=2,3,5,10$ (solid lines) connecting the standard lump solution ($a=1$, dashed-dotted line) with the sine-Gordon kink ($a \rightarrow \infty$,
dashed line).}
\end{center}
\end{figure}

\subsubsection{The energy density}

The energy density for this family of solutions is given by

\begin{eqnarray}
\varepsilon_1(a)&=&\frac{2a}{2^a}\sin^{2a}[\phi_{0}/a]\sin^{2a}[\phi_{0}] \label{energyone} \\
\varepsilon_2(a)&=&\frac{2a}{2^a}\cos^{2a}[\phi_{0}/a]\sin^{2a}[\phi_{0}]. \label{energytwo}
\end{eqnarray}
The evolution of $\varepsilon_{2}(a)$ as we increase $a$ is shown (apart from the factor $2a/2^a$) in Fig. 8, where we see the change of profile from the lump ($a=1$) to the wall ($a\geq 2$) and how the energy density becomes more localized as we increase $a$. $\varepsilon_{1}(a)$ fades out with $a$, vanishing in the large $a$ limit.

\begin{figure}[h]
\begin{center}
\includegraphics[width=8cm,height=4.5cm]{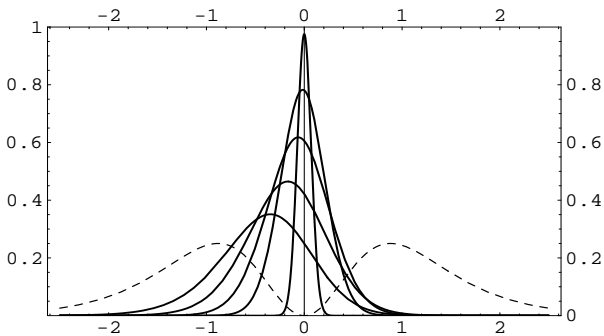}
\caption{\label{fig:3} Behaviour with the distance $x$ of the factor $\cos^{2a}[\phi_{0}/a]\sin^{2a}[\phi_{0}]$ of $\varepsilon_2(q)$ in Eq.(\ref{energytwo}) as (from bottom to top) $a$ evolves
($=2,3,5,10,100$ solid lines),
starting from the canonical kinetic term case ($a=1$, dashed line). Note the change in the shape of the energy density
in the transition from the lump-like solution ($a=1$) to the first wall one ($a=2$).}
\end{center}
\end{figure}

\subsection{Generalizations of families A and B}

For each of the families of models considered before we can defined a more general one including at the same time $V_W$ and $V_L$ for the first
family and $V_1$ and $V_2$ for the second one. To do this we include a constant $C$ into the wall solution $\phi_W$ of the family $A$ as

\begin{equation}
\overline{\phi}=a\sin\left[\frac{\overline{\phi}_0}{a}+C\right], \label{sol-gen}
\end{equation}
in such a way that for $C=0$ we recover the wall solution of (\ref{generalized-lw}) while for $C=\pi/2$ we obtain the lump-like one of (\ref{generalized-lw}). This is indeed a simple rotation of $\pi/2$ of the $\sin$ over the $\cos$ transforming a wall into a lump. Moreover we can write $C=m \pi$ so for $m$ integer the field (\ref{sol-gen}) describes (anti-)walls, while for $m$ semi-integer it describes lumps. Now there is a single potential
containing both families, which is obtained as

\begin{equation}
\overline{V}(\overline{\phi})=\frac{2a-1}{2^a}\cos^{2a}[\arcsin[\overline{\phi}/a]]\cos^{2a}[a(\arcsin[\overline{\phi}/a]+C)].
\end{equation}
The same procedure can be applied to the second family ($B$), obtaining in this case a single field

\begin{equation}
\phi=-a\cos\left[\frac{\phi_0}{a}+C'\right],
\end{equation}
which gives $\phi_1$ of Eq.(\ref{solutions2}) for $C'=0$ and $\phi_2$ for $C'=\pi/2$, transforming one solution into the other through a rotation
of $\pi/2$, as in the former case. The associated single potential is

\begin{eqnarray}
V(\phi)&=&\frac{2a-1}{2^a}\sin^{2a}[\arccos[-\phi/a]] \cdot \nonumber \\ &\cdot& \sin^{2a}[a(\arccos[-\phi/a]+C')].
\end{eqnarray}
Note that in both cases, aside from the cases $C=C'=0$ and $C=C'=\pi/2$, which give the solutions already studied, we have now new solutions for
other values of $C$, which can be interpreted as composed states of kinks/anti-kinks, possibly interacting between themselves, and whose analysis is beyond the aim of this paper.

\section{Projection and deformation methods}

In this section we compare our results with those obtained using a generalization of the deformation method for k-fields.

Let $\varphi$ be generically the cosine/sine-Gordon fields and $V$ the respective potential. Its equation of motion is simply
\begin{equation}
\varphi^\prime = \left(2 V(\varphi)\right)^{\frac{1}{2}},\label{varphi}
\end{equation}

Now let $\chi$ be generically the fields for the k-models $A$ and $B$, and $U$ the respective potential. Its equation of motion is
\begin{equation}
\chi^\prime = \left(\frac{2^a}{2a - 1} U(\chi)\right)^{\frac{1}{2 a}}.\label{chi}
\end{equation}

We shall now apply the deformation method (see Ref. \cite{Bazeia2}) between the sine/cosine-Gordon models and the k-field families,
in both directions.

\subsection{Deforming the k-field models $A$ and $B$ into the canonical cosine-Gordon and sine-Gordon models}

Let us first look for the deformation function that deforms the k-field models $A/B$ into the canonical cosine/sine-Gordon models, respectively.
Thus $\varphi$ will be seen as the deformed field. We also take the deformation function as $f(\varphi)$. Therefore
\begin{equation}
\chi = f(\varphi).
\end{equation}
Using Eqs.(\ref{varphi})-(\ref{chi}) we get that
\begin{equation}
\frac{d f}{d\varphi} = \frac{1}{\sqrt{2}} \Big(\frac{2^a}{2a-1} U(\chi \rightarrow \varphi)\Big)^{\frac{1}{2a}}\,\frac{1}{V(\chi)^{\frac{1}{2}}}. \label{def}
\end{equation}
Making some simple calculations one can obtain the deformation functions associated to the four models defined by the families A/B considered before.
They are given by
\begin{eqnarray}
f^A_L(\varphi)&=&a \cos(\varphi/a)\nonumber\\
f^A_W(\varphi)&=&a \sin(\varphi/a)\nonumber\\
f^B_1(\varphi)&=&-a \cos(\varphi/a)\nonumber\\
f^B_2(\varphi)&=&a \sin(\varphi/a), \label{deformationfunction}
\end{eqnarray}
which exactly reproduce the projection functions given in Eqs.(\ref{generalized-lw}) and (\ref{solutions2}). Through these functions it can be seen that via Eq.(\ref{def}) the potentials of the cosine-Gordon (\ref{potential}) and sine-Gordon (\ref{potential1}) are obtained from the k-field potentials in Eqs.(\ref{modelLa}) and (\ref{modelWa}) for model A and in Eqs.(\ref{model1}) and (\ref{model2}) for model B, respectively. Therefore we conclude that the canonical cosine/sine-Gordon models can be seen as deformations of the k-field models $A/B$, respectively. Note that this procedure transforms a given k-field model (i.e. $a$ fixed) with a finite number of topological sectors, into the periodic (infinite) structure of vacua of the cosine/sine-Gordon system.

\subsection{Deforming the canonical cosine/sine-Gordon models into the k-field models $A$ and $B$}

We now apply the deformation method in the opposite way. Thus $\chi$ will be seen as the deformed field. We now take the deformation function as $g(\chi)$ with
\begin{equation}
\frac{d g}{d\chi}=\frac{1}{\frac{d f}{d\varphi} }
\end{equation}
and one obtains, explicitly
\begin{eqnarray}
g^A_L(\chi)&=&a \arccos(\chi/a)\nonumber\\
g^A_W(\chi)&=&a \arcsin(\chi/a)\nonumber\\
g^B_1(\chi)&=&a \arccos(-\chi/a)\nonumber\\
g^B_2(\chi)&=&a \arcsin(\chi/a),
\end{eqnarray}
which can be shown to transform the cosine/sine-Gordon potentials into the four k-field potentials of models $A/B$ via the deformed potential

\begin{equation}
U(\chi)=\frac{2a-1}{2^{a-1}}\frac{V(\varphi \rightarrow g(\chi))^a}{\left(\frac{dg}{d\chi}\right)^{2a}}.
\end{equation}
Therefore, as expected, we can also conclude that the canonical cosine/sine-Gordon models can be deformed into the k-field models $A/B$, respectively.

\section{Conclusions}

In this paper we have presented a projection method allowing to obtain analytical soliton solutions of wall or lump-like type for k-field theories defined as powers of the standard canonical kinetic term. We have used the sine-Gordon soliton and its cosine-Gordon counterpart, as a system of utmost physical and mathematical interest, in order to illustrate this procedure. The solutions of these k-field models reduce to the standard Klein-Gordon kink and the standard lump in the lowest case ($a=1$) of the projection while some of them approach to the cosine/sine-Gordon models in the limit $a \rightarrow \infty$. During the evolution with $a$ the topology of the solution may change, transforming a lump into a wall or the opposite.

We have shown that this method can be seen as a generalization of the deformation method for k-fields of the form $X^a$. Using this we concluded that the models that we presented, i.e., $A$/$B$  can be seen as deformed models of the canonical cosine/sine-Gordon, respectively, and viceversa. This fact opens the possibility of extending the deformation method to k-field theories with more general non-canonical kinetic terms. This issue deserves a separate analysis, a task to be carried out in a future work.

\begin{center}
\section*{Acknowledgments}
\end{center}

D. R.-G. is partially funded by the Centro de F\'{\i}sica do Porto, and would like to thank the Departamento de F\'{\i}sica e Astronomia da Faculdade de Ciencias da Universidade do Porto for all their hospitality while doing this work. C. dS. is partially funded under the FCT project CERN/FP/116358/2010. The authors would like to thank F. Pacetti for useful discussions. \\

\appendix{\section{Spatially localized and modulated solutions}}

In this annex we present a slight modification of model (\ref{model2}). We define a new lagrangian

\begin{equation}
L_B=X |X|^{a-1}-\frac{(2a-1)}{2^a}\cos^{2a}[b\overline{\phi}_{0}]\cos^{2a}[\overline{\phi}_{0}], \label{breather0}
\end{equation}
\begin{figure}[h]
\begin{center}
\includegraphics[width=8cm,height=4.5cm]{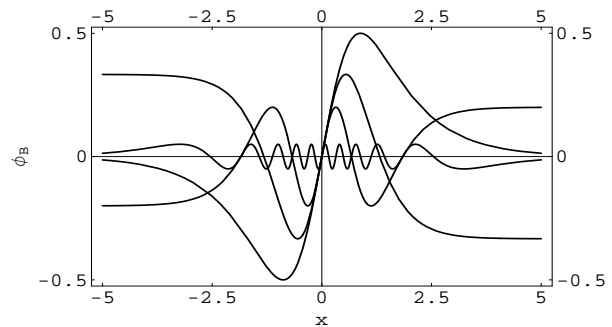}
\caption{\label{fig:3} $\phi_B$ profile in Eq.(\ref{breather0}) versus the distance $x$ for models (\ref{breather1}) for values $b=2,3,5,10$ and $a=2$, showing the
different possible kinds of asymptotic behaviours. Note that the height of the maxima decreases with $b$.}
\end{center}
\end{figure}
\begin{figure}[h]
\begin{center}
\includegraphics[width=8cm,height=4.5cm]{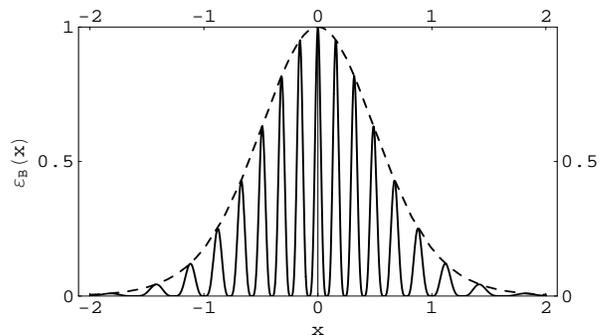}
\caption{\label{fig:3} $\varepsilon_B$ in Eq.(\ref{breather1}) in the case $a=2$ and $b=20$, with the dashed
line representing the envelope $\cos^{4}[\overline{\phi}_{0}]$.}
\end{center}
\end{figure}
where $b$ is a free parameter for a given model. In particular, for the case $b=1/a$ the above model coincides with the one in Eq.(\ref{modelWa}), and its associated solutions with those of (\ref{generalized-lw}). We shall however consider in this annex the case of $b>1$, which describes non-linear solutions which are spatially localized and modulated. The scalar field and the energy density are given by

\begin{eqnarray}
\phi_B &=&\frac{1}{b}\sin[b \cdot \overline{\phi}_{0}]\label{breather0} \\
\varepsilon_B &=&\frac{2a}{2^a}\cos^{2a}[b\overline{\phi}_{0}]\cos^{2a}[\overline{\phi}_{0}]. \label{breather1}
\end{eqnarray}

Note that the profile of this solution does not depend on the power of the kinetic term, but just on the parameter $b$. In figure 9 we have plotted $\phi_B(x)$ for several values of $b$. We see that as $b$ increases the solution shows more peaks while its height decreases. Moreover several distinct behaviours are found as we make $b$ to evolve. Concerning the energy density in Eq.(\ref{breather1}), it describes damped oscillations in space, with an envelope given by $\cos^{2a}[\overline{\phi}_{0}]$, as can be seen in figure 10 for the particular case of $a=2$ and $b=20$.

We believe that this kind of configurations has possible applications in the context of breather solutions, which shall be investigated elsewhere.

\end{document}